# Observation of field emission from GeSn nanoparticles epitaxially grown on silicon nanopillar arrays

*Antonio Di Bartolomeo[1*], Maurizio Passacantando[2], Gang Niu[3*], Viktoria Schlykow[4], Grzegorz Lupina[4], Filippo Giubileo[5] and Thomas Schroeder[4,6]*

[1] Physics Department "E. R. Caianiello", University of Salerno, via Giovanni Paolo II, 84084, Fisciano, Italy

[2] Department of Physical and Chemical Science, University of L'Aquila, via Vetoio, 67100, Coppito, L'Aquila, Italy

[3] Electronic Materials Research Laboratory, Key Laboratory of the Ministry of Education & International Center for Dielectric Research, Xi'an Jiaotong University, Xi'an 710049, China.

[4] IHP Microelectronics, Im Technologiepark 25, 15236 Frankfurt (Oder), Germany

[5] CNR-SPIN Salerno, via Giovanni Paolo II, 84084, Fisciano, Italy

[6] Institute of Physics and Chemistry, BTU Cottbus-Senftenberg, Konrad Zuse Str. 1, 03046 Cottbus, Germany

[*]*E-mail: dibant@fisica.unisa.it*
[*]*E-mail: gangniu@mail.xjtu.edu.cn*

**Abstract**

We apply molecular beam epitaxy to grow GeSn-nanoparticles on top of Si-nanopillars patterned onto p-type Si wafers. We use X-ray photoelectron spectroscopy to confirm a metallic behavior of the nanoparticle surface due to partial Sn segregation as well as the presence of a superficial Ge oxide. We report the observation of stable field emission current from the GeSn-nanoparticles, with turn on field of $65\ V/\mu m$ and field enhancement factor $\beta \sim 100$ at anode-cathode distance of $\sim 0.6\ \mu m$. We prove that field emission can be enhanced by preventing GeSn nanoparticles oxidation or by breaking the oxide layer through electrical stress. Finally, we show that GeSn/p-Si junctions have a rectifying behavior.

**1. Introduction**

The availability of improved nano-patterning techniques and of new nanostructured materials in the last two decades has created renewed interest in vacuum nano-electronics [1-2]. New high-current and long-lifetime electron sources, so-called cold-cathodes, have been introduced. These sources are based on the phenomenon of field emission (FE), which is the injection of electrons from the surface of materials into vacuum by quantum tunneling under the influence of an applied electric field [3]. Cold cathodes find applications in high-power and microwave vacuum electronic devices [4-5], flat panel displays [6], scanning and transmission electron microscopy (SEM and TEM) [7] or in electric propulsion systems [8]. Recent popular candidates as cold-cathode materials are carbon nanotubes [9-13], nanodiamonds [14], graphene [15-16], Si, ZnO, SiC, GaN, AlN semiconducting nanowires [17-21] or metallic nanowires and nanotips [22-24].

In this work, we report the fabrication of germanium nanoparticles alloyed with a small fraction of tin (GeSn-NPs), on which we perform field emission studies. We show that the surface of such GeSn NPs has a metallic behavior due to Sn segregation, while their core is essentially a Ge semiconductor. To our best knowledge, no previous observation of field emission from GeSn-NPs has been reported, despite some published works on field emission from Ge nanowires [25], Ge cone arrays [26], and Ge nanostructured thin films [27].

Here, we measure a stable field emission current, which makes GeSn-NPs considerable for FE applications. Besides, GeSn nanostructures are of particular interest for optoelectronics [28-29], due to their tunable bandgap, the transition from indirect to direct bandgap, the high electron and hole mobility and the compatibility with silicon technology [30-32].

**2. Experimental details**

Figure 1a) displays a schematic layout of the device and the measurement setup used in this work. The device consists of an array of GeSn nanoparticles, grown by molecular beam epitaxy (MBE) on the top of Si nanopillars (Si-NPLs) patterned on a p-type Si wafer. A SEM tilted-view image of the GeSn-NPs array is shown in figure 1b).

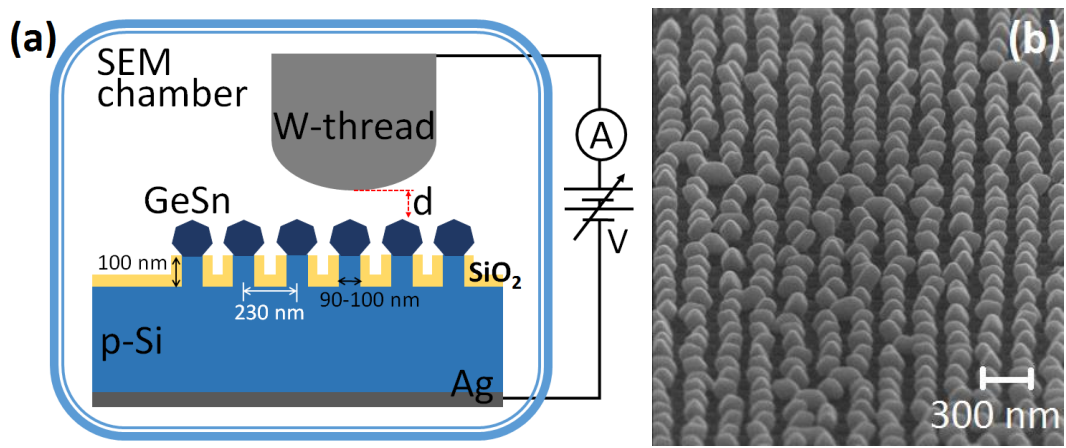

Figure 1. a) Layout of the GeSn-NPs/Si-NPLs device and measurement setup. The device is located in a SEM chamber equipped with a tungsten thread with fine positioning control over the sample. b) SEM image of GeSn-NPs array.

Prior to the epitaxial growth of GeSn-NPs, the patterned wafers were chemically cleaned by a 0.5 at.% HF solution for 12 s followed by a rinsing in deionized water for 1 min. The growth was carried out at a substrate temperature of 750 °C for 20 min by evaporating Ge and Sn species in a DCA MBE chamber with a base pressure of ~2×10$^{-10}$ mbar. Ge was evaporated using an electron beam gun with a growth rate of ~7.0 ± 1.5 nm/min while the Sn was evaporated from a Knudsen cell heated to ~1050 °C, targeting a Sn concentration of max. 4%, according to calibrations using GeSn films grown on Si at lower temperatures [33]. Due to the high growth temperature (750 °C) and the intrinsic solubility limitation of the GeSn alloy, the Sn concentration in GeSn is expected to be less than 1%; furthermore, it is likely that Sn segregates forming nanoclusters on the Ge-NPs. Actually, X-ray diffraction (XRD) out-of-plane (004) diffraction yielded a 0.2% Sn content, which makes nanoparticles core essentially Ge-like.

Figure 2a) is a top view of the Si-NPLs displaying a pitch of 230 nm and a diameter of 90-100 nm. The pillars have height of 100 nm and side-walls and interspaces covered by $SiO_2$. Their detailed fabrication can be found in [34-35]. Figure 2b) shows that GeSn-NPs have a dominant elongated shape with lateral maximum dimension in the 100-200 nm range, while figure 2c), at higher magnification, clearly evidences the mentioned Sn nanoclusters on the surface of GeSn-NPs due to Sn segregation.

To further investigate the chemistry of GeSn-NPs surface, X-ray photoelectron spectroscopy (XPS) was carried out. Figure 2d) shows the XPS core level peaks of Ge and Sn (black line) and the valence band region (red line). No contaminant species are detectable within the sensitivity of the technique. Only a very low peak from adsorbed carbon is present on the spectra. The binding energy (BE) calibration of the spectra has been referred to C 1s peak, located at BE = 284.8 eV. The peak located at about 29 and 32 eV in figure 2d), can be assigned to germanium and germanium oxide, respectively. In particular, from XPS analysis as a function of the take-off angle (not shown here), we can conclude that the core of the NPs is Ge (with negligible Sn content) while the air-exposed portion of the nanoparticles are covered by $GeO_2$. Figure 2d) confirms also the presence of Sn at the surface. The Sn segregated at the surface is likely the responsible reason for the significant amount of electrons shown by the valence band region at the Fermi level.

In general, GeSn-NPs are semiconducting, and the effect of Sn doping is the narrowing of the Ge bandgap and its possible transformation from indirect to direct bandgap [32, 36], which makes them interesting for optoelectronic applications. However, in our case, the content of Sn in the core of the nanoparticles is too low to have any significant effect.

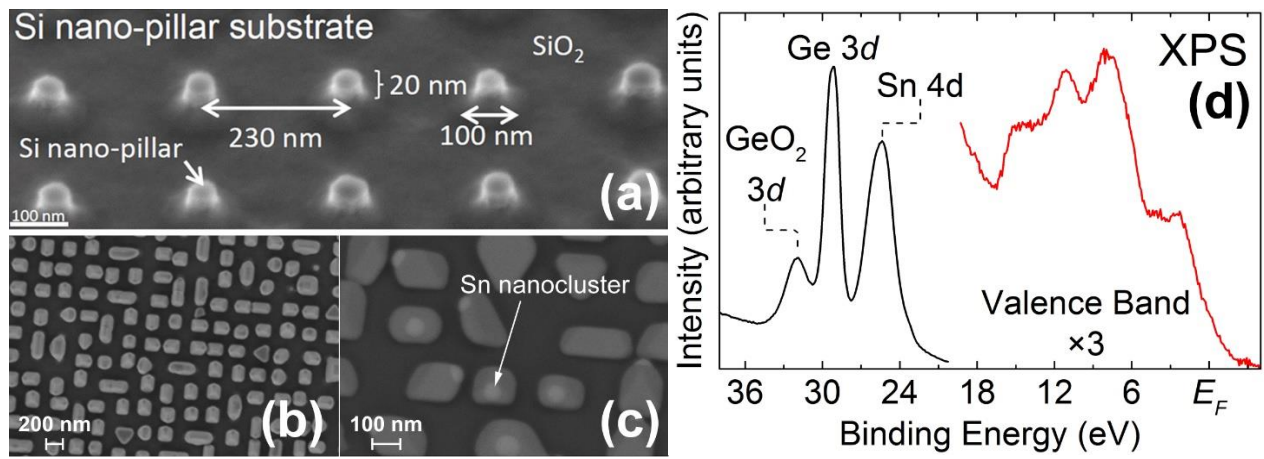

Figure 2. a) SEM top-view of an array of Si-NPLs emerging from $SiO_2$ used as seed for Ge-NPs epitaxial growth. b) Top view of Ge-NPs showing elongated shapes with lateral size in the range 100-200 nm. c) SEM image showing nanocluster of Sn on the GeSn-NPs of figure 2b). d) XPS of the Ge and Sn core levels (black line) and valence band (red line).

Electrical measurements were performed with the sample at low pressure (< $10^{-6}$ mbar) inside a SEM chamber. A tungsten thread (W-thread), of ~4 µm diameter, was used as anode at variable distances from the GeSn-NPs array, while the cathode consisted of a layer of silver painted on the scratched backside of the Si wafer to assure ohmic back-contact. The SEM imaging, combined with nanometric movement control of the W-thread (by Kleindiek manipulators), allowed controlling the W-thread vertical position (indicated as $d$ in figure 1a). I-V measurements were performed using a Keithley 4200 source measurement unit.

**3. Results and discussion**

Figure 3a) shows I-V curves obtained with the W-thread in non-physical contact with the GeSn-NPs array and at decreasing distances from it. For $d < 1\ \mu m$, a current emerges from the setup floor noise, which is $\leq 0.1\ pA$ on the entire sweeping range. This current, initially fairly unstable (blue curve), gets more regular after few sweeps (red curve) and exhibits a steep rise of more than six orders of magnitude over the short range of $\sim 25V$, and reaches a saturation around $0.3\ \mu A$.

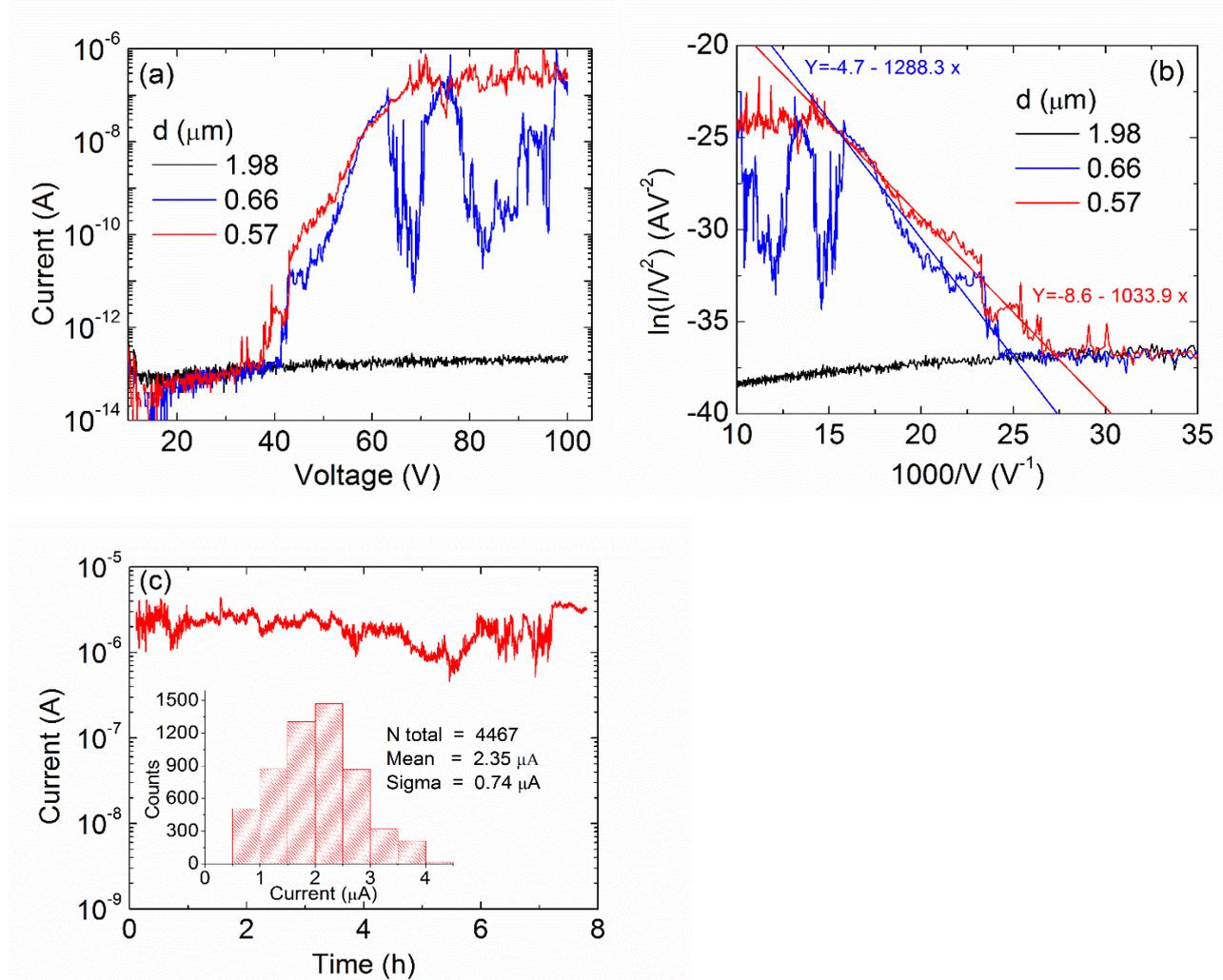

Figure 3. a) Current measured with the W-thread at given distances from the GeSn-NPs array. Successive sweeps result in more stable FE current (red curve). b) Fowler-Nordheim plot of data in a). c) Time stability of FE current from GeSn-NPs (bias 70 V, distance d = 0.50 µm).

This behavior is typical of field emission (FE) current [9-11] and is usually described by the Fowler-Nordheim (F-N) model:

$$I = Sa\frac{E^2}{\phi}exp\left(-\frac{b\phi^{3/2}}{E}\right) \quad (1)$$

where S is the emitting area, E is the local electric field (surface field), $\phi$ is the workfunction of the emitting material and $a$ and $b$ are constants ($a = 1.54 \times 10^{-6} A\ eV\ V^{-2}$ and $= 6.83\ V^{-3/2}\ V\ nm^{-1}$). The saturation is usually attributed to series resistance and/or to space charge limited conduction [9]. Eq. (1) is easily verified by checking the linearity of a plot of $ln(I/V^2)$ vs. $1/V$ (known as F-N plot), which is reported in figure 3b). The electric field between the cathode and anode separated by a distance $d$ with a potential difference $V$ is given by $\tilde{E} = V/d$. The emission turn-on field, here defined as the applied field being able to take the current above the floor noise to the value of 1 pA, is $\tilde{E}_{on} = V/d \approx 65\ V/\mu m$. Due to its sharp form, the local electric field at the surface of a nanoparticle, that is the field that appears in eq. (1), can be higher than $\tilde{E}$ by the so-called field enhancement factor $\beta$:

$$E = \beta\tilde{E} = \beta\frac{V}{d}.$$

Assuming as workfunction of GeSn-NPs the one of Ge, $\phi_{Ge} \approx 5\ eV$ [37], the field enhancement factor can be estimated from the slope of the straight lines in the F-N plot as $\beta \approx 105 - 130$. However, it is likely that the main contribution to FE is from Sn nanoclusters segregated on GeSn-NPs surface, due to their lower workfunction. In such case, being $\phi_{Sn} = 4.42\ eV$, the field enhancement factor would be $\beta \approx 90 - 110$.

The current stability over time is an important ingredient for possible technological applications. For GeSn-NPs, stability was checked and demonstrated over a period of about 8 hours, as shown in figure 3c).

Figure 4a) shows I-V characteristics measured with the W-thread in physical contact with the GeSn-NPs array. The forward sweep curve (black squares) shows a rectifying behavior with reverse saturation current at positive bias. The GeSn-NPs are intrinsic (or slightly n-doped for effect of Sn). i-Ge/p-Si heterojunctions with rectifying behavior and reverse current at positive bias have been previously reported [38]. Hence, we believe that the rectifying behavior originates at the GeSn/Si interface, while W/GeSn junctions form an ohmic contact, consistently with the reported metallic behavior of the GeSn-NPs surface. The W/GeSn-NPs contact resistance can be affected by the presence of a native Ge and Sn oxide layer formed upon exposure of GeSn-NPs to air, as well as to W oxide on the W-thread. An appreciable hysteresis is seen between the forward (first) and backward (second) sweep in figure 4a), with the second sweep corresponding to higher current. Likely, the electrical stress induces defect states in the oxide layer that increase its conductivity. A negligible contribution to the current can be due to GeSn-NPs getting in closer contact with the W-thread for electrostatic attraction.

Figure 4b), c) and d) present the evolution of the I-V characteristics for successive voltage sweeps, that is for increasing electrical stress. Figure 4b) shows the current behavior (full red circles) for physical contact on a pristine area of the GeSn-NPs array: the I-V curve well mimic the ideal reverse current of a diode (GeSn/Si in this case), having a value around 10 nA and weak dependence on the reverse bias. The lower reverse current, as compared to the one of figure 4a), can be attributed to the presence of an unbroken $GeO_x/SnO_x$ layer on the surface of the GeSn-NPs, which adds an extra series resistance. The oxide layer has the twofold effect of limiting the current [39] and the effective voltage applied to the GeSn/Si junctions.

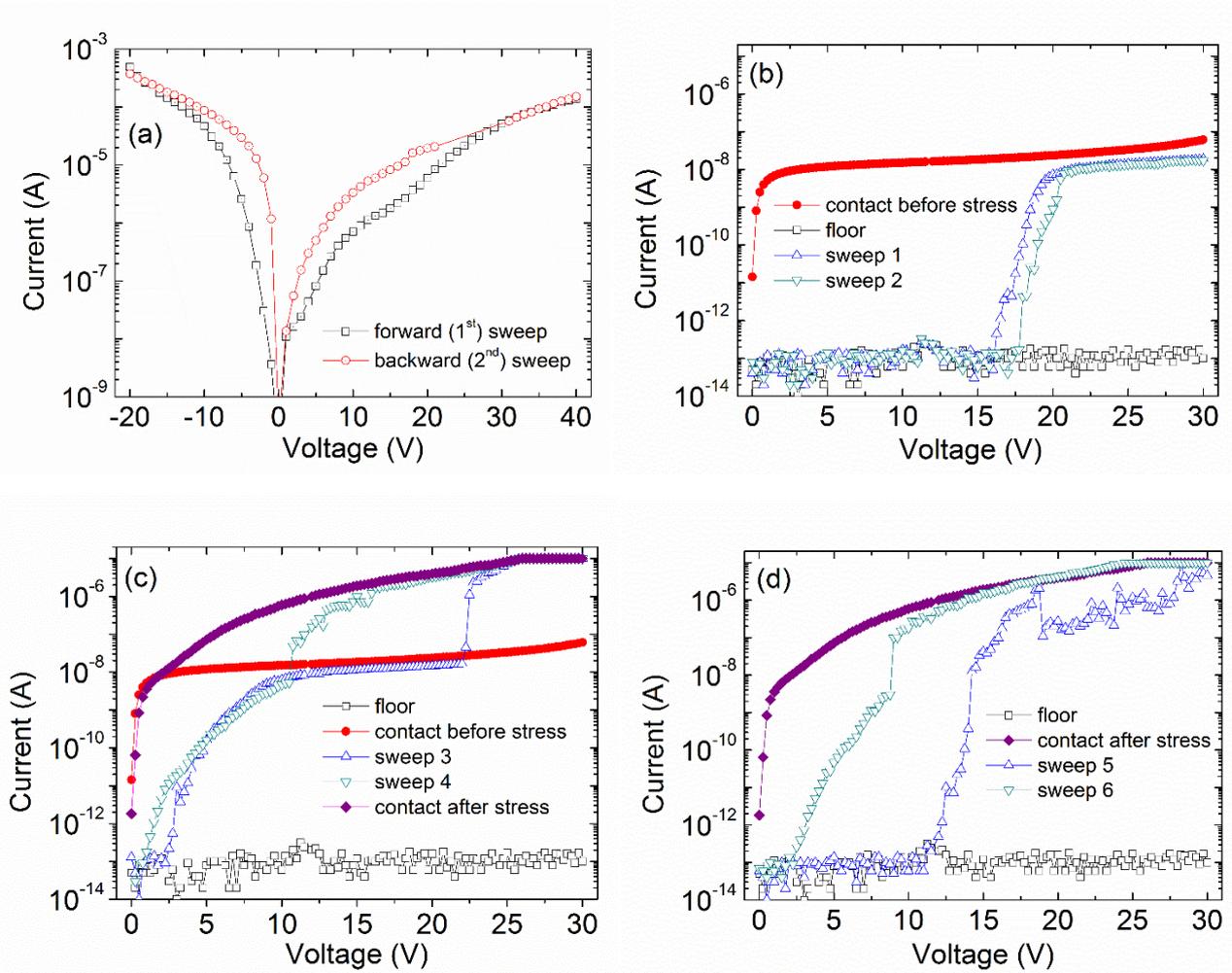

Figure 4. a) I-V characteristics measured with the W-thread in physical contact with the GeSn-NPs array. b) I-V characteristics with the W-thread in physical contact (solid symbols) and at different distances (empty symbols) from the sample. c) I-V curves showing oxide electric breakdown, which results in lower W-thread/GeSn-NPs contact resistance. The red circles and purple diamonds correspond to physical contact before and after electrical stress, respectively, while empty symbol curves correspond to FE and electric breakdown. d) FE current (empty symbols) after $GeO_x/SnO_x$ electric breakdown. Sweep 6 is taken at lower distance than sweep 5.

The other I-V curves in figure 4b) correspond to the W-thread in non-physical contact with the sample. At distances of several hundred nanometers no current is measured (floor), while at lower distances a FE current emerges from the floor and rapidly approaches the limit represented by the contact current. Examples of such current are labelled as sweep 1 and 2 (empty symbols) in figure 4b). With decreasing $d$, the electric field between the W-thread and the GeSn-NPs can overcome the dielectric rigidity of the oxide layer and provoke a dielectric breakdown. An electric breakdown can happen also at the GeSn/Si junctions, which can then lose their rectifying properties. In both cases, breakdown could result in a current higher than the contact one of figure 4b). This is shown in figure 4c, where the current (sweep 3 or 4) rapidly reaches the contact current limit and follows it for a while, until suddenly overcomes it for the appearance of a breakdown at $V \sim 22\ V$ for sweep 3 and $V \sim 11\ V$ for sweep 4, respectively. The electric breakdown of the oxide layer or of the GeSn/Si junctions is confirmed by next sweep (full purple diamond curve) measured with the W-thread in physical contact with the GeSn-NPs underneath. The new maximum contact current still remains 1-2 orders of magnitude lower that the current measured in forward bias, pointing more towards the breakdown of the oxide layer rather than to that of the GeSn/Si junctions. The lower resistance opposed by the broken superficial $GeO_x/SnO_x$ layer also increases the effective reverse voltage applied at the junction, hence making its current more bias dependent. Finally, figure 4d) reports successive I-V characteristics measured after detaching the W-thread from the sample. A FE current with a higher limiting value is observed confirming that a higher FE current from GeSn nanoparticles can be achieved either by preventing oxidation or by applying an electrical stress suitable to break the covering oxide layer.

## 4. Conclusions

In conclusion, we have grown and characterized GeSn-NPs on Si-NPLs arrays and measured a FE current from them. As far as we know, no previous observation of FE from GeSn-NPs has been reported. We have shown that such current has good stability and can be enhanced by preventing the formation of a superficial oxide layer or by breaking it through the application of a suitable electrical stress. As byproduct, we have shown that intrinsic GeSn/p-Si junction forms a rectifying junction and that the GeSn-NPs surface has a metallic surface behavior due to Sn segregation effects.